# Liquid Mirrors: A Review

Ermanno F. Borra, Centre d'Optique, Photonique et Lasers
Département de Physique, Université Laval, Canada G1K 7P4
Fax: 418-656-2040
Email: borra@phy.ulaval.ca

# ABSTRACT

The surface of a spinning liquid takes the shape of a paraboloid that can be used as a reflecting mirror. This very old and nearly forgotten concept has recently been revived and I review its present status. Extensive interferometric tests of liquid mirrors (the largest one having a diameter of 2.5-meters ) show excellent optical qualities. I discuss the factors that can limit the optical quality of liquid mirrors, how to minimize them as well as the basic technology. A handful of liquid mirrors have now been built that are used for scientific work. I show representative data obtained from 2.65-m diameter liquid mirror telescopes used for astronomy and the atmospheric sciences (lidar). Section 5, of particular interest to cosmologists, or astronomers using surveys, examines the expected performance of 4-m liquid mirror telescopes dedicated to cosmological surveys. It is rather impressive, due to the fact that the instruments work full-time on four-year surveys: Spectrophotometry reaches B=24 for all objects within over 100 square degrees and wide-band photometry reaches about B=28. I consider the future of liquid mirror telescopes: limits to their sizes, engineering issues as well as speculations on lunar or space LMTs. I briefly mention the possibility of non-rotating GRIN liquid mirrors. Finally I address the issues of the field accessible to LMTs equipped with novel optical correctors. Optical design work, and some exploratory laboratory work, indicate that a single LMT should be able to access, with excellent images, small regions anywhere inside fields as large as 45 degrees.

## SOMMAIRE

La surface d'un liquide en rotation prend la forme d'une parabole qui peut servir comme miroir réfléchissant. Je passe en revue cet ancien concept, presque oublié, qui a été récemment reconsidéré. Des tests interférométriques de miroirs liquides (le plus grand ayant un diamètre de 2.5 mètres) démontrent des qualités optiques excellentes. Je discute les facteurs qui peuvent limiter les qualités optiques des miroirs liquides, comment les minimiser, ainsi que la technologie de base. Un petit nombre de miroirs liquides ont étés construits et sont utilisés pour des travaux scientifiques. Je montre quelques résultats représentatifs de deux télescopes à miroir liquide ayant des diamètres de 2.65 mètres qui sont utilisés pour l'astronomie et les sciences atmosphériques (lidar). La section 5, particulièrement intéressante pour des cosmologistes ou des astronomes intéresses aux relevés, examine les possibilités de télescopes à miroirs  de 4 mètres de diamètre dédiés à des relevés cosmologiques. Leur rendement est impressionnant, grâce au fait qu'ils travaillent à temps plein sur des relevés de 4 ans de durée. Ils atteignent la 24éme magnitude pour des relevés spectrophotomètriques pour tous les objets contenus dans une bande de ciel dépassant 100 degrés carrés, tandis que de la photométrie à grande bande passante atteint la 28ème magnitude. Je considère le futur des miroirs liquides: les limite à leurs dimensions, des questions de génie, ainsi que des spéculations traitant sur les TML spatiaux et lunaires. Je mentionne la possibilité de miroirs liquides stationnaires utilisant des optiques GRIN. Finalement je discute le sujet des champs accessibles à des TML équipés avec des correcteurs innovateurs. Du design optique, ainsi que quelque travail de laboratoires, indiquent qu'un seul TML devrait être capable d'accéder, avec des excellentes images, à des petites régions n'importe où dans des champs aussi grands que 45 degrés.

## 1. Introduction

It has been known for at least a couple of centuries (see [1] for a historical review) that the surface of a spinning liquid takes the shape of a paraboloid that could, in principle, be used as the primary mirror of a telescope. The first known mention of liquid mirrors was made by the Italian E. Capocci [1]. The concept was never taken seriously for two main reasons. First, early attempts to make such mirrors were only partially successful [2], giving a bad reputation to the concept. Second, liquid mirrors were only considered for astronomical applications for which they have an obvious limitation: They cannot be tilted and therefore cannot be pointed and cannot track like conventional telescopes, a major handicap that made them all but useless to Astronomy. However, as pointed out by Borra [3] modern technology now gives us alternate tracking techniques that render liquid mirrors useful to astronomy: For imagery, narrow-band filter spectroscopy or slitless spectroscopy, one can use a technique, called time delayed integration and abbreviated as TDI, that uses a CCD detector that tracks by electronically stepping its pixels. The information is stored on disk and the nightly observations can be coadded with a computer to give long integration times. The technique has been used for some time with a fixed telescope [4] and imagery with a liquid mirror telescope has been demonstrated by Hickson et al. [5]. High and medium resolution spectroscopy can also be adapted to fixed telescopes. For example, Weedman, Ramsey, Ray, and Sneden [6] are implementing a fiber tracking system, that feeds the light to a fixed spectrograph, with a transit telescope. Indeed Borra [7] has argued that essentially any type of astronomical instruments could be adapted for observations with a fixed telescope.

I was originally drawn to consider liquid mirrors because they promise two main advantages over conventional glass mirrors: They are considerably cheaper and it should be possible to build them to much larger diameters. Having concluded that it would be worthwhile to explore again the liquid mirror concept [3] my team began a feasibility study to determine whether, in practice, it is possible to generate an optical quality surface on a spinning liquid. Early simple tests [8] were encouraging so that we built better testing facilities that showed that a 1.5-m diameter liquid mirror had such good optical quality that it was diffraction-limited ([9], hereafter referred to as Paper I). This was followed by tests of a 2.5-m mirror [10]. Also, to test liquid mirrors in an astronomical environment, we operated 2 simple telescopes of 1-m and 1.2-m diameters and obtained over 200 hours of data on the sky. This has led to a milestone: the first publication describing astronomical research done with a liquid mirror telescope [11]. A collaboration between the University of British Columbia and Laval is now operating a 2.65-m diameter LMT situated near Vancouver [5] to carry out a survey of a strip of sky.

Liquid mirrors are interesting in other areas of science besides Astronomy. For example, atmospheric scientists have expressed great interest for these inexpensive large mirrors for Lidar applications: The University of Western Ontario has built a Lidar facility that houses a 2.65-m diameter liquid mirror as receiver. Liquid mirrors have, or promise, interesting properties for many optical applications: very high surface quality, very low or very high numerical apertures, variable focus that can be controlled with a very high precision.

## 2. Liquid Mirrors

It is straightforward to show [3] that, in a rotating fluid, adding the vectors of the centripetal and gravitational accelerations gives a surface that has the shape of a parabola. Using a reflecting liquid one therefore gets a reflecting parabola that could be used as the primary mirror of a telescope. The focal length of the mirror L is related to the acceleration of gravity g and the angular velocity of the turntable $\omega$ by

$$L = g/(2\omega^2). \quad (1)$$

For large mirrors of practical interest the periods of rotation are of the order of 10 seconds and the linear velocities at the rims of the mirrors range between 5 and 20 km/h. Table 1 shows some of the characteristics of selected mirrors. Figure 1 shows a liquid mirror having a diameter of 2.5 meters and a focal length of 3 meters that we have extensively tested in our laboratory at Laval [10].

Fig. 2 gives an exploded view of the basic mirror setup. The mirror and bearing are fixed to a three-point mount that aligns the axis of rotation parallel to the gravitational field of the Earth. Alignment is done with a spirit level to within one arcsecond, sufficient for many applications. There are optical methods that can align it to a greater accuracy.

We presently use airbearings because they are convenient for small systems and commercially available units have the required precision and low friction. For larger mirrors it will be preferable to use oil-lubricated bearings as they have greater stiffness and can support substantially higher masses for a given bearing size. Given the low rotational velocities involved there is no appreciable gyroscopic effect; hence the mirror wobbles with the bearing. The displacement of the image caused by the wobble is readily evaluated with geometrical optics: If the bearing tilts by an angle $\theta$, the image in the focal plane moves by $2\theta$. Our airbearings have radial and axial errors of 1 micron and peak to valley coning errors < 0.2 arcseconds. For astronomical applications, this is acceptable since the RMS value is well below the average seeing on Earth (>1 arcsecond). Should the wobbling be excessive, one should use a more accurate bearing or, alternatively, simply reduce it with an active mount or a wobble plate [12,13].

The turntable is driven by a synchronous motor coupled to it via pulleys and a thin mylar belt made from a length of magnetic tape obtained from a discarded audiocassette. The motor is controlled by a variable-frequency AC power supply stabilized with a crystal oscillator. We control the rotational velocity of the table and thus the focal length of the mirror with the frequency of the power supply.

We have made containers with a variety of construction techniques, from simple flat plywood disks for 1-m diameter mirrors to light-weight composite material paraboloids for larger ones. Our latest containers are made of Kevlar laminated over a foam core [14, 15]. The final figuring of the top of the containers is done by spincasting a polyurethane resin. Earlier mirrors were spuncast with epoxy but we found that they were sensitive to temperature variations and behaved like bimetallic plates, since the coefficients of thermal expansion of Kevlar and epoxy are very different, warping with temperature variations. Spincasting with a soft urethane resin solves this problem. To spincast, the turntable is spun while we pour the liquid resin in the container. It takes the shape of a parabola and is allowed to harden while the table is spinning.

To minimize weight and therefore cost, and to help dampen disturbances, we have developed techniques that allow us to work with layers of mercury as thin as 1-mm (Paper I). The mercury layer can be thought as a thin liquid high reflectivity coating.

Girard [1] made a detailed analysis of the costs of the components and materials needed to build a complete 2.7-m diameter liquid mirror as well as the time to build and install it. It is based on costs and times spent making and installing a previous system and is summarized in Table 2. We can see that this liquid mirror costs 1 to 2 orders of magnitude less than a conventional glass mirror and its cell. Note that this represents the cost of building a prototype so that a better engineered system would probably be less expensive.

[1] L. Girard. Ph.D. thesis Université Laval, in preparation (1995).

## 3. OPTICAL TESTS

Figure 3 in Paper I shows a block diagram of the testing setup. We work at the center of curvature and therefore must use null lenses to correct the large spherical aberration present at the center of curvature of a parabolic mirror. The interferometry is done with a scatter plate interferometer [16]. The interferograms are captured with 1/60 second exposure times by a 512X480 CCD detector connected to an 8-bit framegrabber interfaced to a microcomputer. They are analyzed with software that uses a Fourier technique [17] capable of giving a substantially greater resolution and signal-to-noise ratio than the usual fringe-following algorithms.

Figure 3 shows a typical interferogram of the 2.5-m mirror and Figure 4 shows a typical three-dimensional rendering of its surface. The statistics associated with it are given in units of surface deviations on the mirror at a wavelength of 6328 Å. The spatial resolution of the interferometry is 3 cm on the mirror. The 1/60 second capture times are sufficiently short that we can detect rapid liquid movements, but they also render the interferometry sensitive to seeing and the effects described at the end of this section. Fortunately, we work in a basement room lined with thick concrete walls that has considerable thermal inertia, hence small temperature gradients, so that seeing effects are minimal. Because the mirror is liquid and can shift shape, a few interferograms are not necessarily representative of its optical quality We have analyzed numerous similar interferograms (e.g. [10]) and have videotaped hours of interferogram data that satisfy us that the interferogram shown in Figure 3, and the wavefront of Figure 4 are representative.

We have observed the Airy-like diffraction pattern of the 2.5-meter diameter liquid mirror. Figure 5 shows an image of the point-like object, created by the laser and a spatial filter and captured with a 1/30 second exposure. Fainter rings are present further away from the center of the PSF but are below detection in this image. We have videotaped hours of data and find that the Airy pattern is always visible, although the intensity and symmetry of the rings vary a little, probably from seeing in the testing tower. The observation of the Airy pattern (taken through the null lenses) does not add quantitative information that is not already given by the interferometry; but it gives an easily understood direct evidence that we are near the diffraction limit.

Our videotapes reveal that the centroid of the PSF moves, describing a curve having a peak to valley amplitude ~ 1/3 arcseconds and an RMS amplitude ~ 0.1 arcseconds. The PSF always follows the same path in the focal plane with a period equal to the period of rotation of the mirror. The amplitude of image motion is compatible with the coning error quoted by the builder of the bearing. The amplitude increases if we decrease the air pressure of the airbearing and if the weight distribution of the mirror is unbalanced. We therefore conclude that it is due to the coning error of the bearing. This slow and periodic image motion is small compared with the median seeing at the best observing sites and, if needed, could be corrected in real time with a wobble mirror or an active mount. It obviously could be reduced with a more precise (and expensive) bearing.

Liquid mirrors are not overly sensitive to vibrations. Our mirrors are located in the basement of a large building that vibrates as it is shaken by the wind, by people walking, elevators running, etc... We do see the effect of vibrations in the form of concentric rings on the surface of the mirror. However, the amplitudes of the rings are very small ( ~/100 $\lambda$) for thin mercury layers (Paper I). Vibrations may be a problem for small mirrors operating in a very noisy environment or in the upper floors of buildings. As repeatedly emphasized in this article, one should work with as thin a layer of liquid as possible to dampen all disturbances.

The focal lengths of our mirrors are stable, although we do find some secondary effects due to an insufficient regulation of the rotational velocities (Paper I). The surface tension of mercury only affects the outer few centimeters of our mirrors. Paper I gives a more detailed discussion of effects that degrade the optical qualities of liquid mirrors.

Although our published (and unpublished) data show respectable Strehl ratios, one may wonder what degrades a mirror's quality, preventing it from being perfect. Seeing in the testing tower is a contributor but we also find that time-varying focus and spherical aberrations are measurable causes of deviation from a perfect parabola. There are measurable improvements in the surface quality statistics when we only consider 95% and 90% of the aperture, indicating a departure from a parabola at the edges. Inspection of the wavefronts indicates that the edges display the periodic warping discussed in Paper I that is caused by a varying rotational velocity. Plots of residual spherical aberration and defocus for individual wavefronts show periodic variations that are well-correlated. Obviously the rotational velocity of the turntable is not perfectly stable, causing a periodically varying focus. Computer simulations show that a focus variation, combined with the null lenses and our reduction procedure introduces residual spherical aberration with a ratio of - 1.5 between the amplitudes of spherical aberration and defocus as seen in our data [18]. A better drive will correct this. We also find variable coma having a small amplitude ($\pm$ 0.2 $\lambda$). Computer simulations show that this amplitude is expected from the 0.1 arcsecond wobble of the turntable. Note that the coma and spherical aberration we are talking about are introduced by the null lenses and are not on the mirror surface, only defocus is.

Our extensive optical tests therefore shows that liquid mirrors work and do reach diffraction limit. We know where the remaining small defects originate and are confident that they can be further decreased with additional effort. We did not bother to do it since the present quality is sufficient for many applications (e.g. Astronomy).

## 4. Technology and practical considerations

Most of the topics discussed in this section are covered in greater details in Paper I. I give below an updated summary along with some new material.

a) handling mercury

Mercury vapor can be detrimental to health [19] if inhaled massively over long periods of time. However, in practice mercury evaporates very slowly so that proper ventilation eliminates all danger. Furthermore, a transparent oxide skin develops in a few hours and very effectively decreases evaporation. Measurements of vapor concentration taken a few centimeters above the surface of a 50 cm mirror (Paper I) detect mercury vapors after starting a freshly cleaned mirror but the concentration decreases after a few hours below 0.05 mg/m3, the legal limit used in most countries that allows a human to work an 8-hour workshift without protective mask. Measurements taken inside a polyethylene enclosure housing our 2.5-m mirror show that mercury vapors become undetectable after a few hours if there is some ventilation of the enclosure. If we cut the ventilation, there is some initial increase of mercury vapors that later slowly decreases with time. Evidently the oxide skin cuts evaporation more efficiently as time passes. The quantities of mercury involved are small (a 3-m mirror needs about 10 liters) and a simple plastic-lined pool can handle catastrophic spills. In practice, the health and environmental impacts of mercury are insignificant, provided simple measures are taken.

Liquid mercury has a reflection coefficient of about 78 % that varies little with wavelength through the visible and infrared. This value is smaller than the one of freshly evaporated aluminum (92% in the visible); but an aluminum coating deteriorates as it oxidizes, collects dirt and is corroded by acid atmospheric pollutants. Aluminizing a large glass mirror is time consuming ( a few days) so that astronomical mirrors are seldom aluminized more often than once a year. Mercury is very easy to clean, it takes about 1 hour to clean a 3-m mirror and several hours for its surface to stabilize so that it can be cleaned often. We have compared the reflectivities of mercury and aluminum mirrors in our laboratory [20] and find that mercury has 90% of the reflectivity of an aluminum coated mirror. The reflectivity of mercury does not vary over several weeks within the accuracy of our measurements ( ~ 3 %). Although throughout this paper I assume that the reflecting liquid is mercury, it should be possible to make liquid mirrors with other reflecting liquids. For example, we have made a gallium alloy 1-m diameter mirror. I have also considered alloys of the alkali metals that could be used in a lunar or space telescope [21, 22].

The surface of a mercury mirror is easily cleaned. After stopping the mirror, we drag on its surface a plastic tube filled with water, collecting all surface contaminants to one side where they are suctioned off, along with some mercury, with a vacuum pump. Dirty mercury is stored and eventually filtered so that no waste is generated. It is filtered by passing it through a glass funnel lined with a paper filter pierced with a few holes, 1 mm in diameter. One must be careful that mercury does not contact other metals, such as copper or aluminum, for it amalgamates them and mercury containing minute quantities of other metals tarnishes very rapidly.

b) Thin layers

This is a crucial technology and we have dedicated considerable effort to find ways to minimize the thickness of the layer of liquid we work with. There are two main advantages to thin layers. First, because disturbances (wind, vibrations) are dampened more effectively. Second from the fact that cost is dominated by the bearing and the container, the cost of which increase with weight and therefore the depth of the liquid.

If we pour mercury in a glass container, surface tensions break it into drops and puddles until there is enough liquid to cover it uniformly with a thickness of about 4 mm, the minimum needed to make a uniform layer. It is a wetting problem, with the contact angle between the liquid and a solid surface given by Young's equation as $\cos\theta = (\gamma_{sv} - \gamma_{sl})/\gamma_{lv}$, where θ is the contact angle and γ represent the surface tensions between the solid, vapor and liquid interfaces. Mercury has a very high surface tension and does not wet most materials, with the notable exception of metals with which it forms amalgams. Unfortunately mercury amalgamated with most metals (e.g. aluminum) oxidizes very rapidly forming an opaque skin.

In principle, wetting can be promoted by treating appropriately the surface of the container (cleanliness, roughness, molecular surface makeup) or by decreasing the surface tension of mercury (purity, surface monomolecular layers, etc...). Wilkinson [23] gives an extensive review of the surface and other properties of mercury. We have tried several techniques to promote wetting by mercury but met little success (Paper I). We suripenditously found that mercury sticks more to polyurethane surfaces than to epoxy ones.

In practice, we discovered after much experimenting a dynamic technique that allows us to start a mirror with a layer as thin as 2 mm. The container needs a groove along its circumference so that the layer is more than 4 millimeters thick in a circular region a few centimeters wide extending along the rim. We pour a quantity of mercury sufficient to fill the container to the desired thickness. The turntable is then spun by hand to a rotational velocity significantly higher than allowed by the parabolic shape of the container, so that all mercury collects in a ring at its periphery. We then gradually slow down the container until it uniformly fills with mercury. This procedure may have to be repeated a few times before meeting success.

We can make very thin layers (< 1 mm) by first making a thick layer of mercury and then removing the unwanted excess liquid (Paper I gives technical details). This may seem pointless as one must first start with a large quantity of mercury that overloads the container and bearing anyway; however, it is quite important since thin layers dampen waves and disturbances.

c) damping of disturbances

The attenuation of a disturbance, having an initial amplitude $A_0$, as a function of distance x can be described by

$$A = A_0 e^{-\alpha x}, \qquad (2)$$

where α is a damping coefficient. For a thin layer of liquid disturbances can be dampened by a variety of mechanisms: dissipation inside the liquid itself, damping due to interaction with the bottom and effects due to surface tension. Non-linearities render an exact treatment difficult and one must use approximations. The damping coefficient in a shallow liquid can be approximated by [24]

$$\alpha = \frac{k^3}{\beta^2 n}\left[\left(\frac{\mathrm{Cosh}(4kh)+\mathrm{Cosh}(2kh)-1}{\mathrm{Cosh}(4kh)-1}\right) + \frac{\beta}{2k}\mathrm{Csch}(2kh) - \cdots\right] \qquad (3)$$

where h is the depth of the liquid layer,

$$\beta = \left[\frac{\omega}{2\nu}\right]^{1/2}, \qquad (4)$$

$$\omega = \frac{2\pi}{P} = \sqrt{gk\,\mathrm{Tanh}(kh)}, \qquad (5)$$

and n, the ratio between group and phase velocity is given by

$$n = \frac{\left[\frac{d\omega}{dk}\right]}{\left(\frac{\omega}{k}\right)} = \frac{1}{2}\left[1 + \frac{2kh}{\mathrm{Sinh}(2kh)}\right] . \qquad (6)$$

For long wavelengths ($\lambda$>> 1 cm) we are dealing with gravity waves for which the main restoring force is due to gravity while for shorter wavelength capillary waves the main restoring force is due to surface tensions. Both cases can be treated by equation (3) with the introduction of an effective gravity given by

$$g_{eff} = g\sqrt{1+\left(\frac{\omega^2 r}{g}\right)^2} + T\,\rho^{-1}\,k^2 \qquad (7)$$

where T is the surface tension and $\rho$ the density of the liquid. The $(\omega^2 r/g)^2$ term takes into account the small contribution to the effective gravity from the centrifugal acceleration. The first term in (3) takes into account internal damping while the second one represents the interaction with the bottom of the container. Equation (3) shows that one should work with as thin a layer of mercury as practical, so that all disturbances are dampened as quickly as possible. This theoretical treatment is only in qualitative agreement with the experimental data and the experimental damping values are always larger than the theoretical ones, by factors as large as 4 [25. 26].

d) wind induced disturbances

The surface of a liquid mirror is sensitive to turbulent winds. Sheltering a liquid mirror from the environment is easily done with a silo. The effects of winds induced by the rotation of the mirror are more worrisome and may eventually limit the size of liquid mirrors. Table 1 shows the linear velocities of the rims of some mirrors and gives a measure of the maximum wind that a mirror feels, but we do not know at what wind speed the mirror will run into trouble. Theoretical computations are difficult since there is no satisfactory theory that quantitatively predicts wind driven water waves ( [27] and references therein). Our situation is further complicated by the rotation of the container.

Fortunately, there are solutions to the wind problem. A transparent plastic cover gives a brute force approach that works but degrades the mirror quality. We have examined samples of numerous plastic films [28] finding that some have surprisingly good quality, a conclusion also reached by Thompson [29]. They do degrade the mirror but the overall quality is still comparable to the one of a typical conventional glass mirror, and quite acceptable for survey work at the best astronomical sites. Transparent films carry a few penalties: they are birefringent, they absorb light (a few percents), reflect it (several percents), and, acting like Fabry-Perot interferometers, modulate energy distributions by a few percents [28].

Monomolecular layers calm wind-driven water waves and we can expect that they will do the same for mercury. While the dampening effect of monomolecular layers on water is an active field of research, less is known for mercury but there is some work that we can use to guide us ([30], and references therein). Her work shows that monolayers spread very easily on clean mercury. We did carry out some limited experiments with monomolecular layers that were unsuccessful (Paper I), probably because of surface contaminants. Thin mercury layers decrease the amplitude of waves and will be effective against wind driven waves.

We have tested a 2.5-m mirror having a 0.85-mm thick layer that has an excellent optical quality. Thinner layer may be feasible although they will at some stage be limited by the surface quality of the container.

e) other reflecting liquids

It is desirable to find a reflecting liquid lighter than mercury to decrease costs, for lighter mirrors need less expensive bearings and containers. We therefore are investigating mirrors that use gallium and its eutectic alloys, that have half the density of mercury. Gallium has a relatively high melting temperature ( 30 degrees C) but this is not an obstacle for our experiments show that it is easy to supercool and very stable in the supercooled state. We have supercooled Ga samples to -27 degrees C [20]. We have made a 1-m diameter gallium-indium mirror. A simple qualitative Ronchi test reveals the signature of a parabola and a reasonable surface quality. The main problem that we are facing with gallium is that it oxidizes almost instantaneously. Gallium oxide is transparent and protects the underlying metal from oxidization; however if the liquid is stirred, as occurs during startup, there forms a thick oxide crust that hopelessly degrades reflectivity and surface quality. We have developed a skimmer, that removes the oxide layer, that we are still perfecting.

It would be highly desirable to find a liquid having low density, as well as high reflectivity and high viscosity. An intriguing possibility is to use a low density high-viscosity, but low-reflectivity liquid, and chemically deposit a reflective metallic coating on it. Chemical deposition of thin metal coatings on solids works [31] but still has to be demonstrated on liquids. It may also be possible to increase reflectivity with a dielectric stack.

This is a very young technology and there is much room for improvement. Clearly our mechanical setups should be better engineered, particularly if one wants to build much larger mirrors. Mercury liquid mirrors work but mercury is not an ideal liquid and efforts should be made to find other reflecting liquids

## 5. Applications

### a) Astronomy

In Astronomy, liquid mirrors promise major advances for deep surveys of the sky and, in particular, cosmological studies. This can be understood by considering that conventional telescopes are very expensive and can only be justified by sharing them among many investigators. Obtaining telescope time is a very competitive process so that only a few astronomers manage to get of the order of 3 nights/year on a 4-m class telescope. Further consider that the average time actually spent observing is of the order of 3.5 hours per night [32], the remaining time being lost to weather, technical problems and overhead (slewing the telescope, acquiring and identifying the field, reading the data, etc..). The unfortunate consequence is that it typically takes a decade to gather enough data for a substantial observing program. Furthermore, observing programs requiring more than a week observing time per year on 4-m telescopes do not fare well with observing committees and are simply not envisioned by most astronomers. On the other hand, inexpensive LMTs can be dedicated to a specific project. Cosmological studies involve the observation of a very large number of very faint sources; hence the need for dedicated telescopes and the advantage of LMTs. The outstanding limitation of liquid mirrors, that they can only observe near the zenith, is not a serious handicap for surveys in general and cosmological surveys in particular.

That the limited field of view of a liquid mirror telescopes is not a serious handicap for cosmology is illustrated by Fig. 6 that shows the number counts of quasars, stars and galaxies /square degree, brighter than a given blue magnitude, at the galactic poles. We can see that a survey to B= 22 has access, in a square degree of sky, to 2000 galaxies, 100 quasars and 2000 stars at the galactic poles. The star counts increase substantially as one approaches the galactic plane while the galaxy and quasar counts decrease somewhat because of galactic extinction; but at 60º degrees from the poles we still have 1300 galaxies/square degree and 50 quasars/square degree. Let us only consider the strip of "extragalactic" sky, having a galactic latitude > 30 degrees since a lower galactic latitude carries penalties of excessive crowding and galactic extinction. Note, however, that the "galactic" portion of the night sky can be used for other studies; neglecting it merely reflects the writer's interest. If we conservatively assume that the optical corrector of the telescope yields good images over a 1 degree field, well within the performance of existing corrector designs, it would access about 100 to 200 square degrees of sky, depending on the latitude of the observatory and, at a latitude of 30 degrees, observe well over 250,000 stars, about 250,000 galaxies and about 11,000 quasars with B<22. For comparison, the center for astrophysics redshift survey [33] has so far observed 15,000 galaxies to redshifts < 0.05 in over 15 years of operation. Also for comparison, the total number of quasars in the latest quasar catalog [34] contains 7,000 objects gathered in 30 years, but is essentially useless for statistical studies since the objects were identified from a variety of search techniques having poorly quantifiable selection effects.

Let us consider the performance of a LMT tracking with a CCD in the TDI mode and carrying out spectrophotometry with interference filters, an efficient technique to determine spectral distributions and redshifts of faint objects [35] We shall assume that it observes through discrete filters centered at wavelength $\lambda$, having widths $w_i(\lambda)$ and peak efficiencies $E_{fi}(\lambda)$. Let $S(\lambda)$ represent the number of photons/second received at the detector from an object, $B(\lambda)$ the photon counts/arcsecond/second from the sky background, $n_d$ the dark counts per pixel, $n_r$ the photoelectron equivalent readout noise per pixel and $n_p$ the number of pixels used to determine the magnitude of the object. Considering that the standard deviation of a measurement of N photons is given by √N, a straightforward error propagation analysis gives the signal to noise ratio of a flux measurement as

$$Signal / Noise = St / \sqrt{St + (2\pi / 4)a^2 Bt + 2n_p n_r^2 + 2n_p n_d t} \ , \qquad (8)$$

where t is the integration time, a the diameter of the aperture, projected on the sky, used to integrate the flux from the object. Note that the read-out noise contribution is treated differently and appears as the square of $n_r$ in (8). The factors of 2 come from the assumption that the sky is evaluated in an area equal to the area of the aperture. There are more sophisticated methods to obtain astronomical magnitudes (e.g. profile fitting) but aperture photometry is nearly as efficient for star-like objects in uncrowded fields and is easier to deal with for our purpose. Lilly, Cowie and Gardner [36] have carried out a deep imaging survey of faint galaxies, discuss the merits of different photometric approaches and adopt aperture photometry with a circular apertures of 3 arcseconds diameter. The number of pixels used to perform the integration is given by

$$n_p = \pi a^2/(4p^2), \qquad (9)$$

where p is the pixel size. If $F(\lambda)$ represents the flux/unit wavelength/unit area/unit time received from the object at wavelength λ above the Earth's atmosphere, d the diameter of the telescope, $E_t(\lambda)$ the efficiency of the telescope and instrument, $Q(\lambda)$ the quantum efficiency of the detector, s the "diameter" of the object (seeing disk for a star or quasar), then

$$S = 0.8\, \pi/4\, d^2 E_t\, Q\, E_f\, w\, a^2/s^2\, F \ , \qquad (10)$$

with the additional condition that

$$a^2/s^2 \leq 1 \ . \qquad (11)$$

The 0.8 factor takes into account the transmission of the atmosphere and

$$B = 0.8\, \pi/4\, d^2\, \pi/4\, a^2 E_t\, Q\, E_f\, w\, F_B, \qquad (12)$$

where $F_B$ is the flux/unit wavelength/unit area / square arcsecond/second received from the sky at wavelength $\lambda$. The wavelength dependencies of the terms have been dropped in these equations to lighten them.

Figure 7 shows the signal to noise ratio expected from a 100 second integration time as a function of blue magnitude for a f/1.9 2.65-m diameter LMT (like the UBC-Laval LMT described below) observing through a 200 Å filter centered at 4400 Å. I have assumed a 2048X2048 CCD detector having 15-micron pixels and a read-out noise of 5 electrons (like the CCD of the UBC-Laval LMT). With this detector, 100 seconds correspond to a single TDI nightly pass for a site at a latitude of 32 degrees. I use an aperture of 3.0 arcseconds, as in [36], an object "diameter" of 3.0 arcseconds and a sky background of 22.9 magnitudes/square arcseconds at the zenith. This assumes a dark site and good seeing conditions. Note that the seeing must include the image deformation introduced by the TDI [37]; but at 32˚ latitude the effect is small, since the maximum deviation from linearity at the edge of the CCD is only 0.5 arcseconds.

Photometry of faint objects is not easy, mostly because of the difficulty of accurately estimating the sky background, an effect which is difficult to quantify and is not taken into account in the above derivations. Most of the problem comes from the difficulty of flat-fielding CCD detectors. Fortunately, flat fielding is considerably easier for CCD detectors used in the TDI mode since pixels are moved over an entire column, averaging out irregularities and fringing. Flat fielding is thus done only over one row, easing the problem. As a matter of fact, very accurate photometry of faint objects is often done by driftscanning, a variant of TDI, even with conventional telescopes, precisely to alleviate the flat-fielding problem.

Figure 7 predicts a signal to noise ratio of 10 at B= 21.6 and 5 at B = 22.3. Hickson, Gibson and Callaghan [35] find, from Montecarlo computer simulations, that, at a signal to noise ratio of 10, cross-correlation techniques give rms redshift errors of 6,000 km/sec from low-resolution spectrophotometry for all galaxy types, and morphological type errors less than 0.14, with better performance for early-type galaxies. Similar computer simulations by Cabanac [38] show that the break-finding algorithm discussed by Borra and Brousseau [39] measures 60% of the redshifts of ellipticals and early type spirals to better than 1000 km/sec. This accuracy is sufficient for cosmological studies, in particular studies of the large scale structure of the Universe. One may be skeptical of this kind of redshift accuracy quoted for low resolution and signal to noise ratio spectra; however Beauchemin and Borra [40]successfully detected redshift peaks corresponding to large scale structure found independently by others. This was accomplished with a somewhat higher resolution (75Å) but with noisy photographic spectra.

If we consider sky limited observations ($S \ll Ba^2$), assume that read-out noise and dark counts are negligible, as is the case for modern CCDs and filters as wide as ours, equations (8) to (12) show that, for constant s, a and $F_B$, the limiting flux for a given signal to noise ratio

$$F_l \propto 1/\sqrt{twd^2}\,, \qquad (13)$$

allowing to estimate the performance for longer integration times, wider filters and larger telescopes.

I have so far assumed the LMT telescope that we have constructed and the instrumentation that we have used with it. Let us now conservatively extrapolate by small amounts and consider a 4-m telescope having the same focal length (5-m) so that we keep the same scale at the detector. Let us also assume that we are using a mosaic of 4 CCDs, thus doubling the nightly exposure times and sky coverage; 2X2 arrays have been constructed and larger ones are being planned. The gains in counts and limiting magnitudes computed from (13) are summarized in Table 3, where we also give gains for several nights of observation, in the same filter and for a 1,000 Å filter. Finally we have the gain for an array of four 4-m LMTs.

We can cover the wavelength region from 4000 Å to 10,000 Å with 40 interference filters having logarithmically increasing widths and adequate overlap. The extragalactic strip of sky is observable for about 6 months/year, half of which is "dark time" during which we can observe with 20 of the shorter wavelength filters, the red ones observing during the moon lit nights since the moon’s contribution is less important in the red, given the strong contribution of the upper atmosphere in the red. Assuming 10 hour nights and a good site with only 20% loss due to weather and technical problems, we have 60 dark extragalactic nights/year giving 3 passes/filter/year. In 4 years we would have 12 passes. Table 4 shows that 4 years of observing would get us down to almost 24th magnitude with a S/N = 10, sufficient to get reasonable redshifts, and over 24th magnitude with S/N =5, sufficient for rough redshifts and reasonable estimates of the energy distribution. The increase in sky brightness with wavelength is roughly compensated by the flux increase with wavelength for most faint galaxies, at least for $\lambda < 7{,}000$ Å. We can thus see that we reach faint magnitudes and can address a multitude of unsettled problems such as the nature of the population of galaxies at B = 24 , where an excess of counts occurs and where colors are significantly different from those of galaxies in the local universe [36]. In the same table, I have computed the performance expected for a 1,000 Å filter. It reaches extremely faint magnitudes where a multitude of extragalactic research could be addressed (e.g. supernovae rates at $z>0.4$).

The magnitudes reached are comparable to those reached in the faintest observations done so far [36]. These observations were painstakingly gathered over 3 years on a 3.6-m telescope but their sky coverage (a few square arcminutes) and statistics (about 100 objects) only allowed them a brief glimpse at the universe at those magnitudes and pale with respect to the almost 100 square degrees coverage and millions of objects that a 4 year survey would give. One can get a vivid impression of what the data will look like by examining the photograph of a 200X200 arcseconds CCD frame that reached 28th magnitude in [41]. The LMT survey would get similar data over an area over 100 square degrees, giving a huge database that could be used for a multitude of scientific investigations. Also note that the limiting magnitudes quoted in Table 4 are in good agreement with the 13 hour exposure quoted in [41].

What lessons can we gather from Table 3? *The main one is that the major gain is not obtained by going from a 2.65-m telescope to a 4-m telescope; it is obtained from integration times and that is where liquid mirrors shine.* O course, the same performance could be obtained with a glass mirror but only LMs make it practical since their low cost allows one to afford a telescope dedicated to a specific project. In the same vein, going from a 4-m to a 10-m would only increase the limiting magnitude by 1 magnitude if we can keep the same scale at the detector, difficult to do with present CCDs. In this respect, an array of 4-meters would be preferable for detector matching would be easier. It thus appears that, for survey work in the TDI mode, a 4-m diameter is large enough. If we want better performance, it is preferable to build an array of 4-meters.

A huge increase in performance should also come from a new generations of three-dimensional detectors that measure the energy of the incoming photons as well as their positions [42, 43]; although the technology is still in its infancy. The prospect of carrying out spectrophotometry to B~27 is an exciting one. Note that at least one of theses systems is capable of extremely high spectral resolution [43].

To see how these observations would sample the Universe, I have computed the redshift distribution expected from a survey having lower limiting magnitude $m_0$ and upper limiting magnitude $m_1$ from the usual cosmological relation

$$\frac{dN}{d\Omega dz} = \int_{m_0}^{m_1} \Phi(M) \frac{dV}{dz} dm \quad , \qquad (14)$$

where $\Phi(M)$ is the differential luminosity function (per unit magnitude) of galaxies, $M(H_0, q_0, z, m)$ the absolute magnitude, m the apparent magnitude, $\Omega$ the surface area and $dV(H_0, q_0, z)$ the cosmological volume element, $H_0$ is the present epoch Hubble constant, $q_o$ the deceleration parameter and z the redshift. I use $H_0 = 100$ and $q_0=1/2$. The models have been computed with the mix of galaxy types and the K-corrections described in [44] with the difference that the parameters of the Schechter luminosity function are derived from the CFA survey [33]. I neglect evolution, a reasonable assumption for the redshift depths involved. Figure 8 shows the redshift distributions expected for surveys reaching 22nd and 24th blue magnitudes. A survey to B=28 would reach z > 1.0.

Galaxies are extended objects so that some light is lost by carrying out photometry with a fixed small aperture. The so-called aperture correction is not easy to quantify since it depends on the type of galaxies, redshift, cosmological model, wavelength, etc... A thorough discussion of this difficult problem is beyond the scope of this article but I will simply point out that Lilly, Cowie and Gardner [36] estimate that a 3-arcsecond diameter aperture encloses 95% of light for stellar objects and about 80 % even for their largest galaxies. They also find that the deconvolved galaxy images have 50% light within a 1 arcsecond radius, suggesting that the aperture correction is small even for galaxies having B=21. We shall neglect aperture corrections.

A discussion of the astronomical research that can be done with this data is obviously beyond the scope of this paper but the huge database given by 100 square degrees of images similar to those in [41] can be use for a multitude of projects. For example, a low-resolution spectroscopic survey to B= 24 would sample the Universe to redshifts greater than 0.6 and would observe over a million galaxies in 4 years of observation. By comparison the CFA survey has observed in 15 years 15,000 galaxies to redshifts <0.05. Of course the CFA survey gives redshifts with a precision an order of magnitude better than ours and about 1/3 of our galaxies would be spirals to which we cannot assign reliable redshifts. However, as the CFA team has emphasized several times, the structure in the CFA survey is comparable to the size of the survey and there may be larger structure at greater redshifts.

To demonstrate the use of LMTs as cosmological tools, a collaboration between the University of British Columbia and Laval has led to the construction of a 2.65-m diameter liquid mirror having a focal length of 5 meters [5]. It is equipped with a CCD detector having 2048X2048 pixels. Operating in the TDI mode, every hour of observation produces a strip of sky 20 arcminutes wide and 10 degrees long with 2 minutes integration time. An 8-hour night of observation therefore gives images within 26 square degrees, an area equivalent to the area covered by 132 full moons. The telescope is equipped with a set of 40 narrow-band filters having sufficient overlap to cover a spectral range from 4000 to 10,000 Å. Figure 9 shows an image obtained through a 300 Å filter centered at 7,000 Å. We can see numerous stars and galaxies in this 20 arcminutes frame. We routinely obtain images having FWHM= 2 arcseconds as expected from seeing for a site located near sea level. Although the telescope is located near a large city (Vancouver, B.C., Canada), hence suffers from a relatively bright sky, the 2 minute exposure reveals stellar objects to a red magnitude of R =21. Because the sky is bright, and we do not know his brightness, we cannot directly compare this performance to Fig. 7. However, making a rough estimate of the sky brightness shows a good performance. After having observed for 3 months in the winter of 1994, the telescope will now be moved to a darker site. The main purposes of the project are to gain experience operating LMTs prior to building larger instruments and to study the large scale structure of the Universe with galaxies and quasars.

b) atmospheric science

Atmospheric scientists have recognized that LMTs would allow a significant increase in the power-aperture product of LIght Detection And Ranging (lidar) systems. For example, the region between 30 - 80 km (hence the middle atmosphere) is the least studied part of Earth's atmosphere. This is because the atmosphere is too thin for balloons to float in but too thick for spacecrafts to orbit in. Rockets can fly in it but only for brief periods and therefore cannot carry out long-term studies. Furthermore optical emissions are too weak for passive remote sensing, and the structure of the layers is difficult to detect by radar. On the other hand, this region can be studied by a ground based lidar system having a powerful laser and a large mirror to collect the return light. Most existing lidar systems do not observe off-axis, hence would not suffer from the requirement that the mirror observe the zenith.

An example of the benefit of this increased power-aperture product is the University of Western Ontario's purple crow lidar, designed and built by a group led by Professor R. Sica [2]. Measurements of temperature and density fluctuations in the middle atmosphere have been limited in temporal-spatial resolution by the power-aperture product of current lidar systems. The purple crow lidar combines high-power transmitters with a large diameter liquid mirror. The purple crow lidar is presently capable of receiving Rayleigh-scatter returns from a Nd:YAG laser system and sodium resonance-fluorescence returns from an amplified narrow-bandwith ring dye laser. The Rayleigh scatter experiment measures temperature in the stratosphere and mesosphere, while the sodium system measures the temperature in the mesosphere and lower thermosphere. The receiver is a 2.65m diameter liquid mirror which has been in near continuous operation for 2 years. Several tests show that the mirror behaves like a conventional glass mirror of the same size and that the overall lidar performance is approximately that predicted by the lidar equation (Figure 10) . The lidar system is quite robust, due to the reliable performance of both the laser and the liquid mirror in a wide range of environmental conditions. The combination of a powerful transmitter and a large-aperture receiver allows atmospheric fluctuations due to gravity waves to be studied at extremely high temporal and spatial scales.

[2] R. J. Sica, S. Sargoytchev, E. Borra, L. Girard, S. Argall, C. T. Sparrow, and S. Flatt, In preparation (1994).

Figure 10 compares a Rayleigh-scatter lidar profile measured for a fifteen minute period at 120m height resolution to the profile predicted from the lidar equation, calculated using the measured system parameters and a model atmosphere. The largest uncertainty in the calculation is the atmospheric transmission at 532nm between the ground and 30km, which was assumed to be 60% for this calculation. The measured photocount profile is within 15% of the lidar equation calculation, well within the uncertainties of the calculation. The close agreement of this and other tests conducted by the lidar group at The University of Western Ontario prove that their 2.65-m diameter liquid mirror has equivalent performance to a conventional glass mirror.

c) other applications

At the NASA Lyndon B. Johnson center in Houston, a team led by D. Potter has built a 3-m diameter LMT that has seen first light and will be used to observe for space debris. At the Centre Spatial de Liége, N. Ninane has built a 1.4-m diameter LM that she will use as a reference surface to test a space mirror.

## 6. Field of View

One of the often cited limitations of liquid mirrors addresses the small regions of sky that they can observe. However, this criticism implicitly assumes the corrector designs presently used in astronomical telescopes. Spurred by an article by Richardson and Morbey ([45] I have explored analytically the fundamental limits within which one can correct the aberrations of a liquid mirror observing at a large angle from the zenith [46].

I assumed that correction is applied onto an image of the pupil of the primary mirror, as is done with adaptive optics; then if a telescope observes at $\theta$ degrees from the zenith, assuming perfect correction at the center of the field, the wavefront aberration $\Delta\theta$ arcseconds away is given, to an excellent approximation for $\Delta\theta$ of the order of a few arcseconds, by

$$\Omega(r, \alpha, \theta + \Delta\theta) = \partial W(r, \alpha, \theta)/\partial\theta \, \Delta\theta \quad , \qquad (17)$$

where $W(r, \alpha, \theta)$ represents the wavefront, and r and $\alpha$ are the polar coordinates on the mirror. The surprising result of this simple analysis is that the aberrations can, in principle, be corrected in small patches to zenith distances as high as 45 degrees.

Correcting the pupil is not the best strategy to obtain the widest field of view: I assumed it because a practical limit can be analytically computed. There are known corrector designs, for small $\theta$, that greatly outperform the predictions of equation (17) and presumably correctors exist that outperform them at large $\theta$. The results in [46] should only be used as guidelines with the understanding that better corrector designs are probably feasible.

Following this theoretical exploration we are now investigating practical corrector designs. We designed a simple system consisting of a single active spherical secondary to which were added aberrations up to 7th order. It gave subarcsecond images in small subregions anywhere within a 20 degree field [47], a performance insufficient for imagery, but useful for high-resolution spectroscopy of compact objects. Adding adaptive optics would render it competitive for high-resolution imagery since the field of view would then be limited by the size of the isoplanatic patch. We now have carried out experimental work to determine whether it is possible to mechanically bend a metallic mirror to add several aberrations to its surface. We successfully added third and fifth - order aberrations to a 20-cm diameter mirror [3].

The one-mirror corrector gives adequate performance for spectroscopy but has insufficient image quality and field of view for imagery. We have thus begun investigating with optical design software a family of practical 2-mirror correctors that give far better performance for imaging [48]. We have added an aspheric shape to the conic surfaces of the secondary and tertiary mirrors of the well-known Paul-Baker design [49] and introduced additional degrees of freedom by allowing decentering and tilting of the secondary and tertiary mirrors. We obtain secondary and tertiary mirrors that have reasonable diameters by using only the off-axis segments that actually collect the light rays. The design, dubbed BMW, is versatile and gives a variety of configurations having different focal lengths and geometries. An interesting configuration places the tertiary on the optical axis of the primary so that the tertiary-detector assembly looks like a small altazimuth telescope that views the secondary mirror. To view a different region of sky, the secondary would move on a polar coordinates mount while the tertiary would move in an altazimuth mount. In principle the telescope could track by moving and warping the mirrors to follow an object in the sky, but the shapes of the mirrors would have to change in real time. A simpler system can use rigid mirrors with a survey telescope tracking electronically with a CCD detector. The optical and mechanical setups are then simple since the corrector is set for a particular zenith distance and does not have to be adjusted to work at different zenith distances. A corrector designed for a given zenith distance $\theta$ (e.g. 10 degrees) can be used to observe objects passing anywhere within a field of view of $2\theta$ (e.g. 20 degrees) by moving it at different azimuths but at a fixed zenith distance, as explained by a figure in [48]. Figure 11 [4] shows the spot diagrams for point sources observed at 22.4 , 22.5 and 22.6 degrees from the zenith as well as 3 spots displaced by 0.1 degrees from those in the orthogonal direction. The 6 spots are given by the same BMW corrector of a 4-m LMT.

[3] G. Moretto, T. Bactivelane, M. Wang, M. Ferrari, S. Mazzanti, B. Dibiagio, G. Lemaitre, and E.F.Borra. In preparation (1994 ).

We only have begun exploring designs for correctors working very much off-axis. This is an entirely new area of optical design, since such a need never arose in the era of tiltable telescopes, so that there probably are better designs than those that we have found so far. Once a sufficiently large accessible field is achieved, a fixed primary yields a more efficient system than a classical tiltable telescope. A classical telescope can only observe a field at a time, while a fixed primary with several correctors could access many widely separated fields simultaneously.

## 7. The future

a) How large can one make them?

The promise of gigantic sizes has always been the lure of liquid mirrors; although the discussion in the previous section shows that telescopes in the 4-m class are very competitive and arrays of 4-m could be used for greater collecting areas. We have made so far mirrors as large as 2.7-m, an interesting size but still far from the 30-m diameters originally envisioned [3]. Let us therefore see, in the light of our present experience, what may eventually limit the sizes of liquid mirrors.

[4] G. Moretto Ph.D. thesis Université Laval, in preparation (1995).

The analysis leading to equation (1) assumes a constant gravitational field on a flat earth devoid of rotation; therefore neglecting the curvature of the earth, the Coriolis force introduced by the rotation of the earth, the tides induced by the moon, the effect of the wind induced by the rotation of the mirror. etc... Borra, Beauchemin, and Lalande [27] and Gibson and Hickson [50] have considered the consequences of neglecting these effects. Their main conclusions are that the curvature of the earth introduces a small and easily correctable defocus and that the effect of the tides is negligible. The Coriolis force seems prima facie to be more bothersome since it causes measurable effects. However, the main aberrations predicted are coma and astigmatism that can be eliminated with a small additional correction in the corrector. In practice, however, the values computed in these two articles are upper limits and the effect will be smaller than predicted since they both neglect viscosity and assume that the liquid follows the equipotential surface. Astigmatism and coma lack cylindrical symmetry and, if caused by the Coriolis force, are fixed in the reference frame of the earth, therefore inducing a traveling wave on the surface of the rotating container. These very long waves will be dampened by the thin layers of mercury that we use. Paper I illustrates the effectiveness of thin layers at dampening long wavelength surface wave: A wave having a peak to valley amplitude of 2500 Å with a 5.5 mm layer becomes unobservable (P-V $< \lambda/20$) with a 1.4-mm layer.

b) Engineering issues

The mirrors that we have constructed so far are not particularly well engineered since the emphasis was placed on getting results rather than sound engineering. Throughout our work, a number of engineering issues have arisen that shall have to be addressed prior to building large systems. We have, in collaboration with C. Gosselin of the Laval engineering department and students (e.g. [14]) examined some of them. As mentioned earlier, oilbearings are probably preferable for large system since they have greater stiffness for a given size and therefore can support much larger loads. The stiffness of the container itself is important for the flexure of a centrally supported container, such as pictured in Fig. 2, causes some liquid to flow to the parts that have flexed, increasing the local thickness of liquid, hence the local weight, hence inducing further flexure. If the container is too flexible, this becomes unstable and leads to a catastrophic spill of the liquid.

Arrien [14] has carried out finite element analyses of the containers of liquid mirrors made of a Kevlar skin laminated with epoxy over a foam core (see also [15]). It appears that a centrally supported composite container (Kevlar laminated over foam) should be practical up to diameters of 4-m and perhaps as large as 6-m. However, a larger mirror may have to be supported by an annular bearing to minimize the deflections of the container. A space frame will also help the overall rigidity of the system.

The response of the container under temperature variations is an important issue, given that one should work with a layer thickness of the order of a millimeter. The surface of the mirror should not deviate from a parabola by more than a fraction of a millimeter.

The mirror is also subject to a tilting instability for a small tilt causes liquid to move to one side, resulting in an asymmetric loading that can run away. It can readily be shown [51] that the system becomes unstable if

$$64\, C/(\pi\, \rho\, g\, D^4) > 1\ , \qquad (15)$$

where $\rho$ is the density, D the diameter of the mirror and C the constant of elasticity of the system defined by $M=C\theta$ so that the system tilts by an angle $\theta$ if the moment applied is M . The constant of elasticity C is a function of the elasticity constants of the various components (bearing, container, mount, etc...) and is given by $1/C= \Sigma 1/C_i$. The stability of the system must be checked during the installation of the mirror but it is done once and for all. This is done by adding weights on one side of the mirror and measuring the resulting deflection with a dial gage. We measure the deflection $h_0$ on the side of the weight as well as on the opposite side $h_{180}$. The constant of elasticity is then such that

$$(h_0+h_{180})/2 > R^2W/C > h_{180} , \qquad (16)$$

where R is the radius of the mirror, and W is the test weight. We found that the interface between the container and the bearing is a source of flexure, greatly decreasing C if the surfaces are not well mated. Note that the analysis leading to (15) and (16) assumes that the deformation of the container can be approximated by a plane.

An intriguing engineering solution has been proposed by Vasil'ev [52] who devised an ingenious technique that uses an intermediate damping liquid (IDL). With the IDL technique, a floating container rotates on an intermediate liquid which acts as a damper. The IDL technique gives an inexpensive solution to both the bearing and container problems since it does not require an expensive bearing and, furthermore, the intermediate liquid supports the container over its entire surface. Vasil'ev [52] successfully used it to make a 50 cm diameter mirror but its reflecting surface was an oil and he did not try mercury. We experimented with mercury but found that the container was unstable and spilled. We proposed (Paper I) a solution based on a binary mixture of mercury (for stability and flotation) and a viscous liquid (for damping). We have not made any experiments since.

The main conclusion to derive from this brief discussion is that systems larger than 4 to 6-m should be properly engineered. Engineers tell me that a 30-m mirror is a challenge but that it can be done.

b) Lunar and orbiting LMTs

The astronomical community is considering a new generation of telescopes that are not earth-based. At a conference on the next generation space telescopes held at the Space Telescope Science Institute [53], lunar based as well as space telescopes were considered. A consensus emerged that a 10 to 16-m diameter telescope, either space or moon based should be build, given the enormous advantages that such instruments would have over earth-based telescopes.

Borra [21] has discussed the advantages and feasibility of a large lunar liquid mirror telescope, concluding that the high optical quality and simplicity of a liquid mirror, its low shipping mass, ease of assembly and low maintenance make it a very attractive alternative to a glass mirror.

A space liquid mirror telescope would have several advantages over a lunar-based one. For example, it is less expensive to put a payload in orbit than it is to put it on the moon and a space telescope is subjected to smaller accelerations allowing for a lighter structure. However, liquid mirrors present us with a serious challenge since they need a continuous steady acceleration which, prima facie, seems impractical to obtain in free space. Fortunately, our sun gives us an inexhaustible supply of energy that could be harvested with a solar sail to give the necessary acceleration. Borra [22] has considered the characteristics of very large(10-m to 1000-m diameters) orbiting liquid mirror telescopes (OLMT) continuously accelerated by the radiation pressure from the sun on solar sails. If the velocity of the craft is lower than the right orbital speed, the solar sail would give just enough gravitational acceleration to keep the telescope in orbit. It actually could replace all gravitation to give a stationary instrument capable of long integration times. A rich variety of orbits are feasible with solar sails and orbit switching is possible thus allowing to point the telescope [54] .

The concept may appear to belong more to science fiction than to science and the article by Borra [22] certainly contains speculations and extrapolations of present day technology; however it rests on reasonable assumptions. A feasibility study carried out by NASA in the late seventies [55] has shown that the concept is practical.

A different approach has been proposed by Ragazzoni, Marchetti and Claudi [56] who suggest an OLMT that consists of a liquid mirror warped with magnetic fields. They did not, however, work out the details

The choice of liquid metals for a lunar telescope has been discussed in [21, 57] where it is concluded that mercury is not a good choice, because of high mass and high evaporation rates. A lunar or a space LMT could use low melting temperature gallium alloys (such as gallium-indium-tin) or perhaps alkali alloys, such as sodium-potassium-cesium, that are light and have low melting temperatures. They have good reflection coefficients.

The case presented in [22] is suggestive enough that liquid mirrors should be considered as candidates and further research and development carried out. An alkali 10-m OLMT and its sail would have a mass of about 20 tons [22], less than twice the mass of Hubble Space Telescope and would be within the capacity of present-day launchers so that it could be launched within a reasonable time. Telescopes having diameters from 100-m to 1000-m may be built in the next century, although a 1-km mirror sounds a bit utopian. Liquid mirrors may be the best hope that we have to build a large (4 to 8 meter diameters) lunar telescope in the near future.

Perhaps the most important contribution of Borra [21, 22] and Ragazzoni, Marchetti and Claudi [56] is to point out that there may be ways to make gigantic space or lunar telescopes if one considers entirely novel approaches. Because it will take a long time from the moment the concept is proposed to the moment the instrument will see first light, one should consider radically new technologies very early.

c) liquid GRIN mirror

A new type of non-rotating liquid mirror based on GRIN (gradient index) optics has been proposed by Borra [58] The mirror consists of a stationary container having a thin layer of mercury covered with a transparent liquid in which one introduces a radial dependence of the index of refraction. This system focuses light like a mirror, albeit with the chromatic aberrations of a lens. The index gradient could be achieved with a chemical composition gradient set up by a diffusion-based system using membranes. If working systems can be built, this type of mirror should have essentially no size limit and may be used to build optical telescopes having truly gigantic dimensions (hundreds of meter diameters). Verge and Borra [59] found that it would be possible to obtain a satisfactory mirror with existing chemicals; however, they also found that a binary mixture of liquids is unstable against turbulent convection, rendering the mirror useless. They propose to use a ternary liquid mixture to eliminate the density gradient that drives the convection. We have not carried more work on this subject simply because of a lack of time and resources.

e) Magnetic liquid mirrors

Shuter and Whitehead [60] have proposed to use magnetic fields to shape the parabolic primary of a ferrofluid mercury telescope into a sphere. Having a spherical primary is advantageous for wide-field coverage since a sphere looks like a sphere from all angles. For this reason, the primary mirrors of the telescopes that have very wide fields (e.g. Schmidt telescopes) have spherical primaries. They argue that a 10X10 degree sky coverage is possible. Ragazzoni and Marchetti [61] also have proposed to use magnetic fields to shape flat mercury mirrors to be used as active optics as well as rotating mirrors. They also have proposed to use magnetic fields to shape large space LMTs [56].

## 8. Conclusion

Interferometric tests of liquid mirrors having diameters as large as 2.5 meters show excellent optical qualities. Although the idea is very old, the technology did not exist until modern work began a decade ago. We had to painstakingly develop it and understand the factors that can limit the optical quality of liquid mirrors. This task was made more challenging by the fact that only inexpensive solutions are admissible since liquid mirrors make sense only if they are considerably cheaper than conventional glass mirrors. Although the job is not finished and there is much room for improvement of this very young technology, we do have a working design that is sufficiently robust to be useful for scientific use. For this purpose, we also have an adequate understanding of the behavior of a liquid mirror under perturbations. In other words, we know how to make liquid mirrors that work but one can do better. Our present technology will probably be outdated in a few years but the work done so far, and our identification of problems and glimpses of solutions, should lead the way to future improvements. An example of this is our work on thin mercury layers and our proposal to use other reflecting liquids.

A handful of liquid mirrors have now been built and are used for scientific work: astronomy, atmospheric sciences, space sciences, optical shop tests. More applications are certainly forthcoming given the advantages of liquid mirrors, foremost of which is cost.

Estimates of the performance of liquid mirror telescopes dedicated to cosmological surveys shows that major breakthroughs can be brought in astronomy by using inexpensive LMTs dedicated to a particular project. They do not come from the fact that the mirrors are liquid but rather from the fact that they are cheap and thus affordable. *A very* *important consideration is that the work proposed could be done soon, since there is little extrapolation from the largest LMT that has been built.*

The issue of the field accessible to a LMT is a very important one since the usefulness of a LMT increases with its accessible field. Optical design work, indicates that a single LMT should be able to access fields as large as 45 degrees. We only are at the beginning of this exploratory work and additional effort can probably find simpler systems with improved performance. On the other hand, those correctors can only increase the cost of a LMT and they can only be practical if the total instrument is sufficiently cheaper than a conventional telescope.

Considering the future of liquid mirror telescopes, the most exciting one concerns the possibility of having lunar or space LMTs, in particular the science-fiction-like orbiting LMT having a 1-km diameter. Should it ever be built, it will be a vision to behold; with its 50-km wide solar sail it should be visible from Earth with the naked eye. A non-rotating liquid GRIN mirror is also exciting since it promises gigantic sizes; but a working prototype remains to be built.

Looking back at a decade of work by myself, my graduate students, postdoctoral fellows and other researchers, there is the satisfaction of seeing a laboratory curiosity coming of age and being used, along with the frustration of the passage of time. Looking at the LM presently being tested in our lab, I cannot help but think that I am looking at a miracle: an optical quality surface of liquid mercury 2.5-m in diameter and only 0.5-mm thick.

**Aknowledgments**

The liquid mirror work at Laval has been financed by the Natural Sciences and Engineering Research Council of Canada, by the FCAR program of the province of Québec as well as a grant from the government of Québec (Collaboration Franco-Québecoise). I wish to thank Dr. R. Sica for discussions regarding the lidar. I dedicate this article to the memory of the late Prof. Louis-Marie Tremblay who partecipated in the liquid mirror project [9,11]: He was a good man.

FIGURE CAPTIONS

Figure 1

Liquid mirror having a diameter of 2.5 meters and a focal length of 3 meters.

Figure 2

Exploded view of the basic setup of a liquid mirror.

Figure 3

Typical interferogram of the 2.5-m mirror shown in figure 1.

Figure 4

Three-dimensional rendering of the surface of the 2.5-m mirror. The statistics associated with it are given in units of surface deviations on the mirror at a wavelength of 6328 Å. See the text for an explanation of the remaining defects.

Figure 5

Image of the point-like object, created by the laser and a spatial filter, imaged by the 2.5-m mirror and captured with a 1/30 second exposure. The elongation of the core is due to a ghost image caused by rapid low-amplitude vibrations of the CCD mount.

Figure 6

Number counts of quasars, stars and galaxies /square degrees, brighter than a given blue magnitude, at the galactic poles.

Figure 7

Signal to noise ratio expected from 100 second integration time (a single nightly pass) as a function of blue magnitude for a f/1.9 2.65-m diameter LMT observing through a 200 Å filter centered at 4400Å. I assume aperture photometry with a 3 arcsecond diameter aperture. Table 3 gives offsets for larger telescopes, wider filters and longer integration times.

Figure 8

Redshift distributions expected for surveys reaching 22nd and 24th blue magnitude. A survey to B=28 would reach z > 1.0.

Figure 9

Image obtained through a 300 Å filter centered at 7,000 Å with the 2.65-m diameter UBC-Laval LMT. The exposure is the result of a single nightly pass (140 seconds exposure). We can see numerous stars and galaxies in this 20X20 arcminutes frame taken at a galactic latitude of about 50 degrees (R.A = 15:29:58, DEC= 49deg 3min 30sec epoch =1994.5) . North is at the top and East to the right (from [5]).

Figure 10

Measured Rayleigh-scatter lidar profile (solid line) for a fifteen minute period around 2245 solar local time on August 21, 1994 at 120m height resolution. The dotted line shows the profile predicted from the lidar equation, calculated using the measured system parameters and a model atmosphere. The data was obtained with a 2.65-m diameter liquid mirror. The figure was provided by R. Sica.

Figure 11

Spot diagrams for point sources observed at 22.4 , 22.5 and 22.6 degrees from the zenith with a LMT equipped with a BMW corrector [53] as well as 3 spots displaced by 0.1 degrees from those in the orthogonal direction. The 6 spots are given by the same BMW corrector.

Table 1. Characteristics of selected liquid mirrors grouped by f/ratio

| Diameter (m) | Period of rotation (sec) | Vrim (km/hour) |
|---|---|---|
| | f/2 | |
| 2.5 | 6.3 | 4.5 |
| 5 | 9.0 | 6 |
| 10 | 12.6 | 9 |
| 30 | 22 | 15 |
| | f/1 | |
| 2.5 | 4.5 | 6 |
| 5 | 6.3 | 9 |
| 10 | 9.9 | 13 |
| 30 | 15.7 | 2 |

Table 2. Costs of components and labor needed to construct a 2.7-m mirror.

(adapted from [16]).

============================================================

| Item | Cost of components (1994 $US) | Labor (hours) |
| --- | --- | --- |
| Complete mirror[1] | 15,400 | 290 |
| Safety equipment[2] | 3,800 | 140 |
| Installation[3] | 500 | 80 |
| Total | 19,700 [4] | 510 |

[1] Complete system, including base, mercury, motor, etc... but does not include compressor ( about $3,000)

[2] Includes mercury sniffer, safety brakes and anti-spill wheels

[3] In-situ installation, including balancing, debugging, checking image quality

[4] The cost estimates given in [5] include labor

Table 3. Performance improvements over the 2.65-m LMT described in section 5

| item | Flux Increase | Δm |
|---|---|---|
| 2X2 CCD mosaic | 2 | 0.37 |
| 4-m mirror | 2.27 | 0.45 |
| 3 nights | 3 | 0.60 |
| 12 nights[1] | 12 | 1.35 |
| 60 nights[1] | 60 | 2.22 |
| 240 nights[1] | 240 | 2.98 |
| 1000 Å | 5 | 0.86 |
| 4 X4-m LMTs | 4 | 0.75 |

[1] See text for explanation

Table 4. Limiting magnitudes for a 4-m LMT with a mosaic of 4 CCDs and for an array of 4 4-m LMTs

==============================================================

| | 200 Å filter | |
|---|---|---|
| | S/N= 10 | S/N =5 |
| 1 night | 22.4 | 23.1 |
| 1 year, 3 passes | 23.0 | 23.7 |
| 4 years, 12 passes | 23.74 | 24.44 |
| 4X4-m LMTs, 4 years, 12 passes | 24.5 | 25.2 |
| | 1000 Å filter | |
| | S/N = 5 | |
| 1 night | 23.96 | |
| 1 year, 60 nights | 26.18 | |
| 4 years, 240 nights | 26.94 | |
| 4X4-m LMTs, 4 years | 27.7 | |

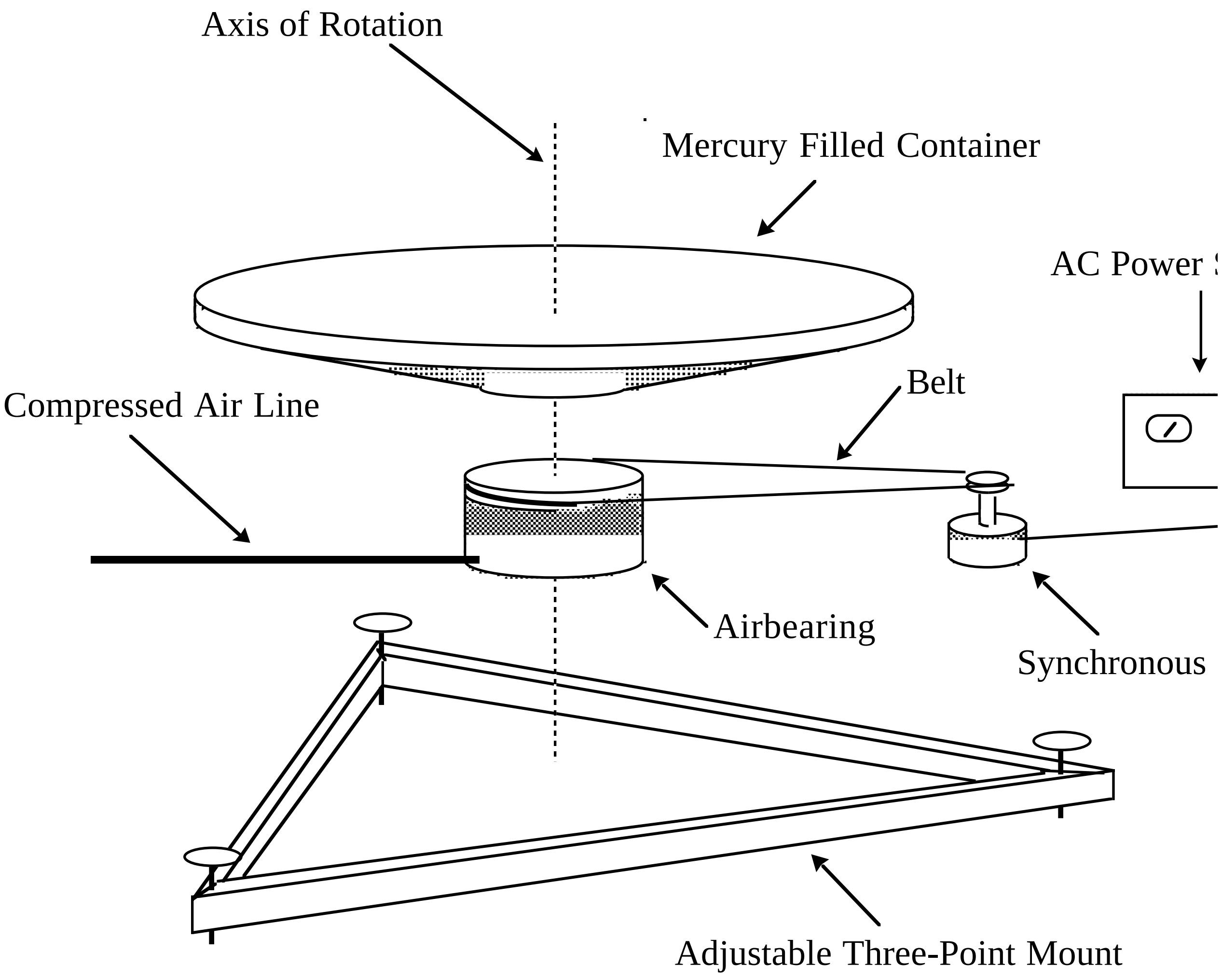

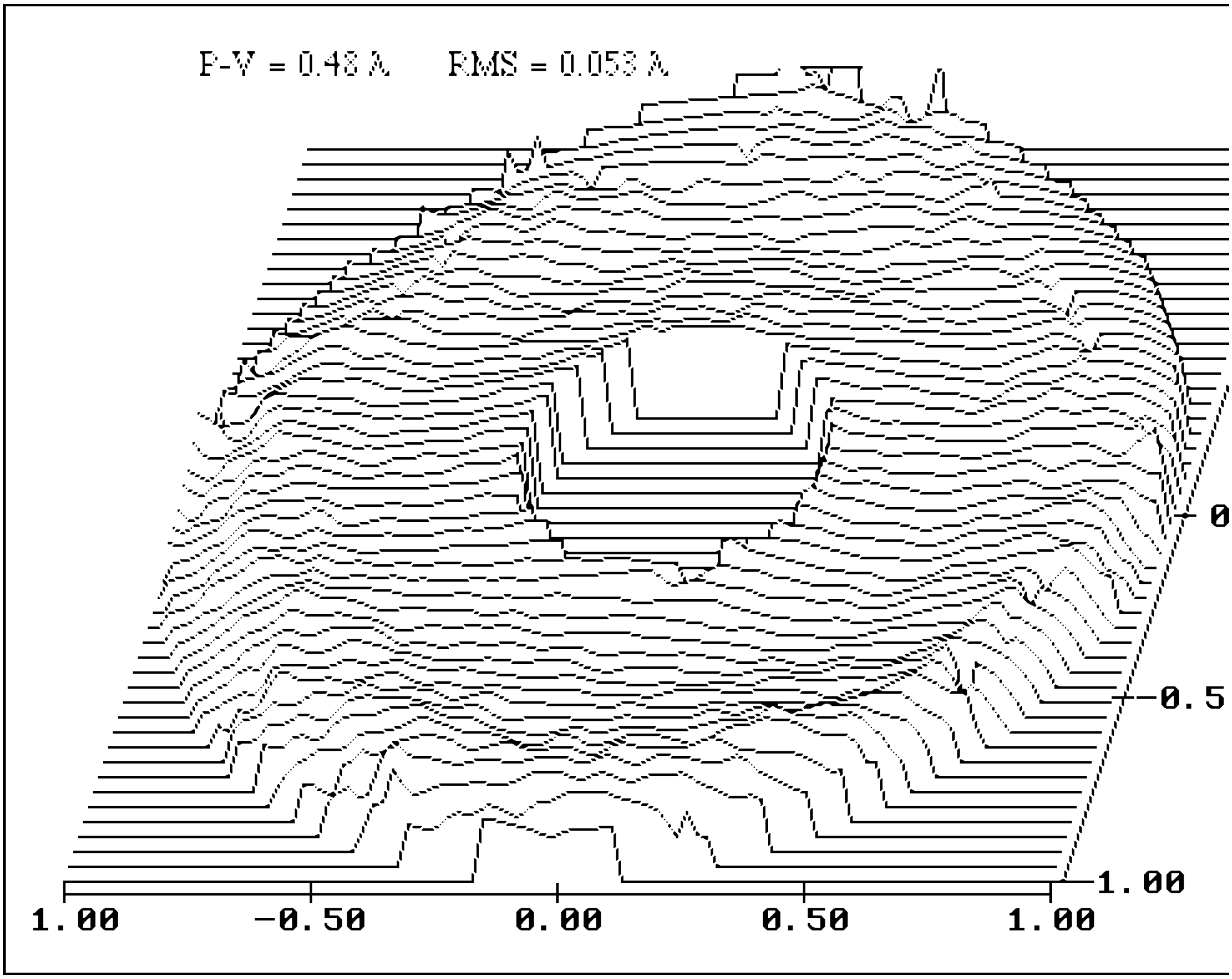

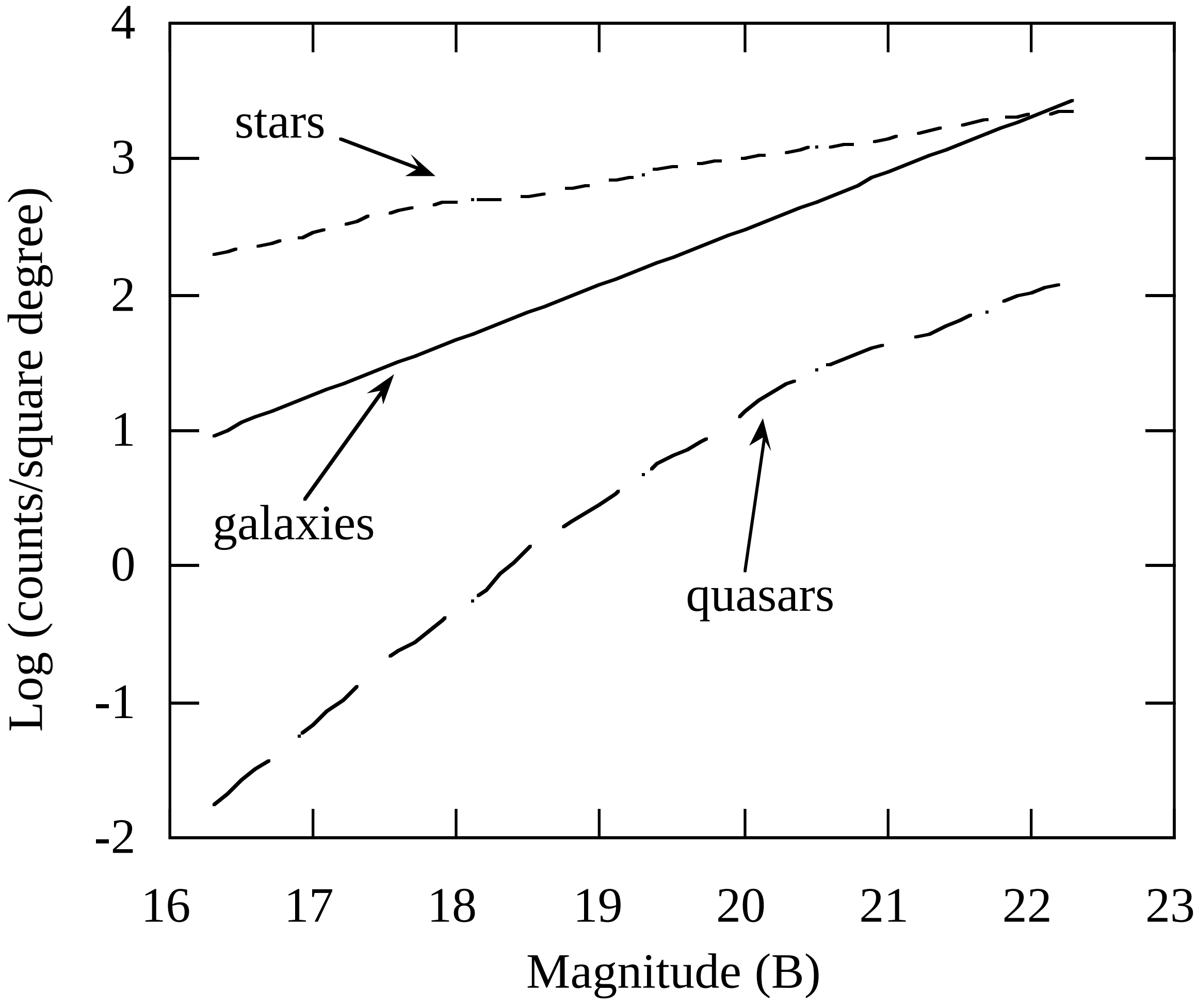

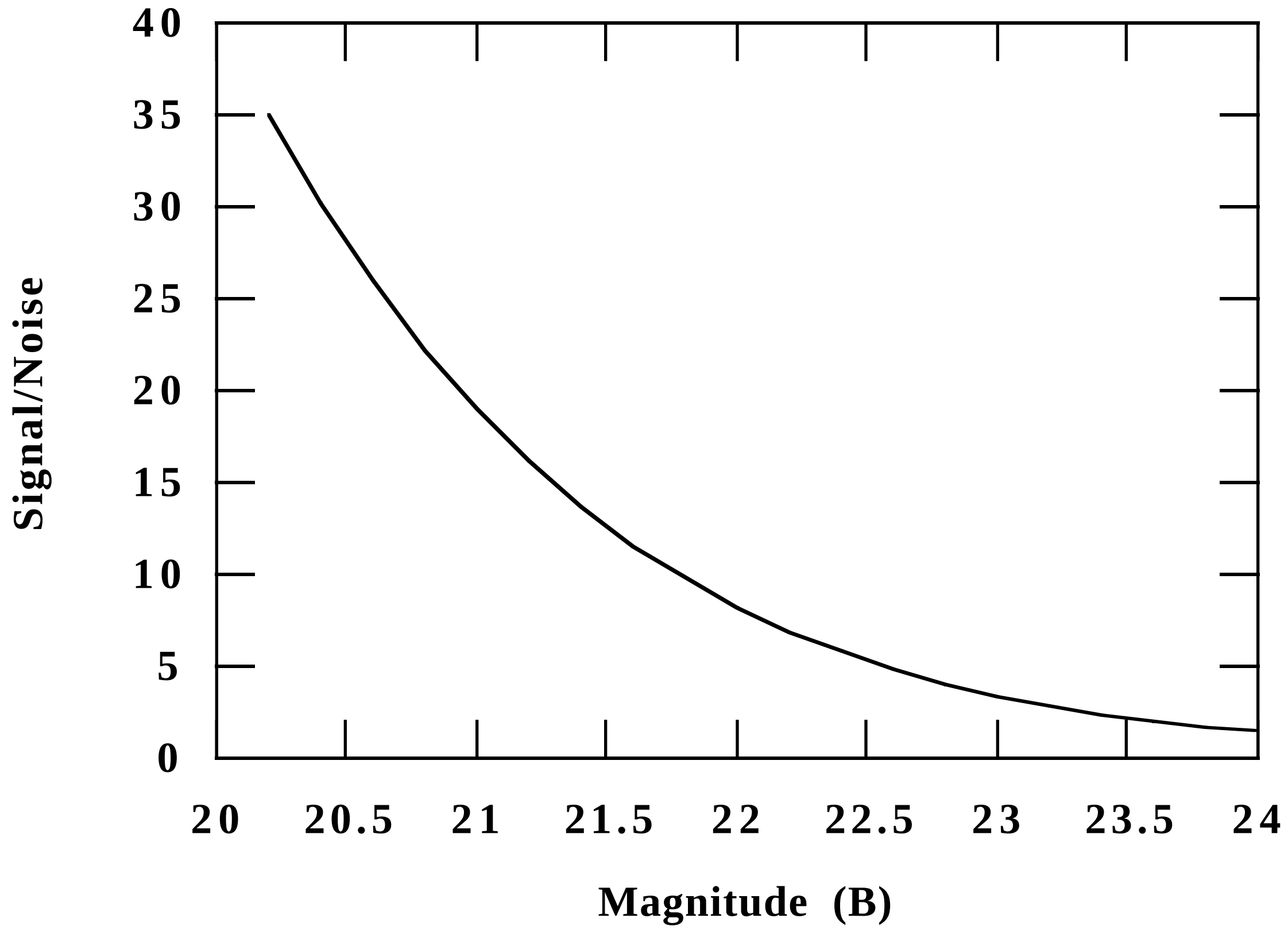

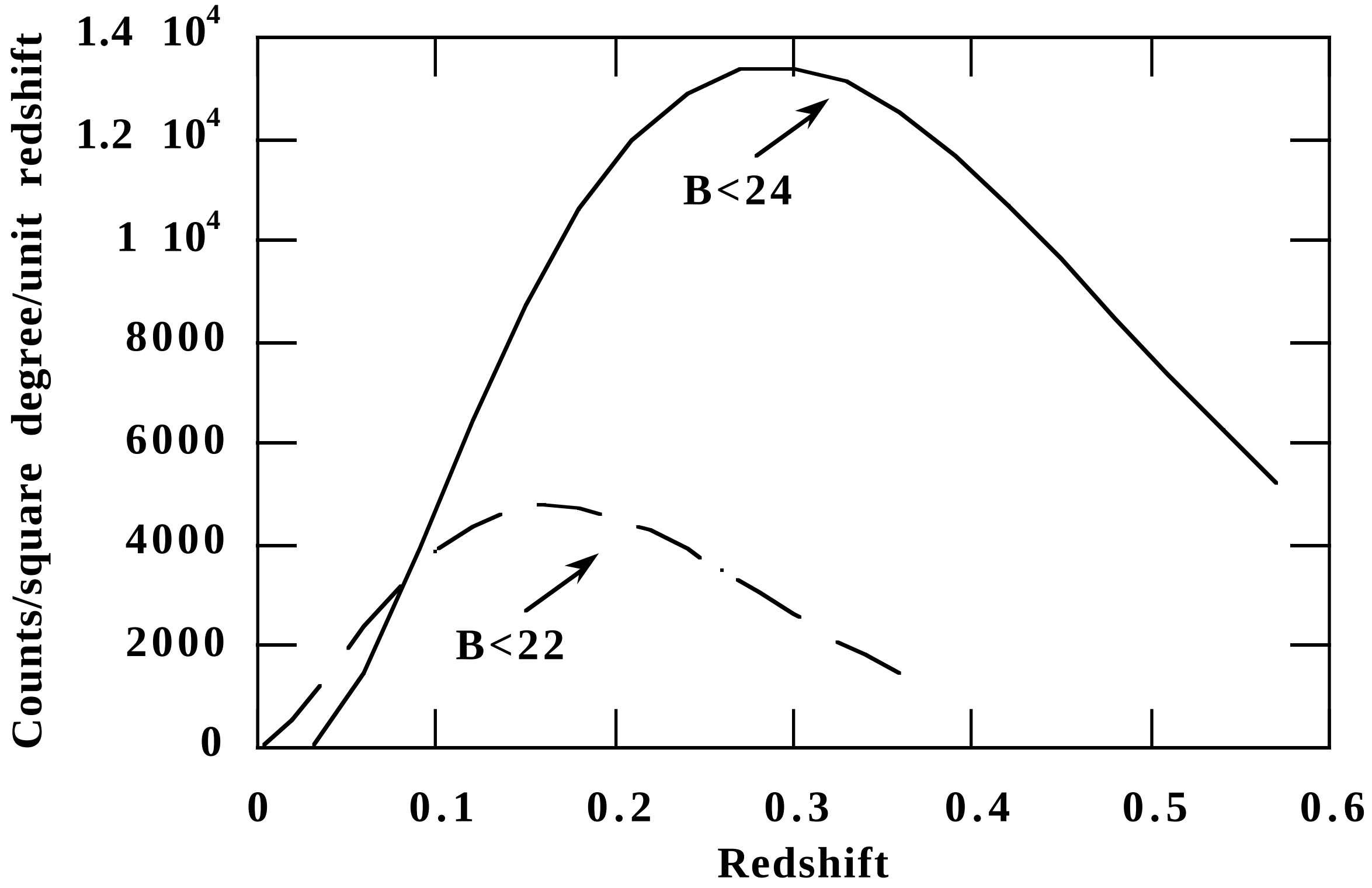